\documentclass[12pt]{iopart}

\usepackage[]{amssymb}
\usepackage[]{amsfonts}
\usepackage[]{latexsym}
\usepackage[]{float}
\usepackage[]{xy}
\usepackage{graphicx}

\newcommand{\mb}[1]{\mbox{\boldmath $#1$}}

\begin{document}

\title[]{
Quantized Berry Phases
\\
for a Local Characterization 
of  Spin Liquids\\
in Frustrated Spin Systems
\footnote{ 
submitted for the proceeding of the conference, 
Highly Frustrated Magnetism 2006 (HFM2006), (June 29, 2006) \hfil\break
http://www.kobe-u.ac.jp/hfm2006/ \hfill
}
}

\author{Y Hatsugai}

\address{
Department of Applied Physics, University of Tokyo
}
\ead{hatsugai@pothos.t.u-tokyo.ac.jp}
\begin{abstract}
Recently 
by using quantized Berry phases,
a prescription for a local characterization of {\em gapped}
topological insulators
 is given\cite{Hatsugai06a}.
One requires the ground state is gapped and is invariant 
under some anti-unitary operation.
A spin liquid which is realized as  a unique ground state
of the Heisenberg spin system with frustrations is
a typical target system,
since  pairwise exchange couplings are always time-reversal  invariants 
even with frustrations.

As for a generic 
 Heisenberg model with a finite excitation gap, 
we locally modify the Hamiltonian by a continuous $SU(2)$ twist only at a specific link
and define the Berry connection 
 by the derivative.
Then the Berry phase evaluated by the entire many-spin wavefunction 
is used to define the local topological order parameter at the link.
We numerically  apply this scheme 
 for several spin liquids and show its physical validity.
\end{abstract}

\maketitle

\section{Topological Orders}

In a modern condensed matter physics, 
a concept of the  symmetry breaking has  a fundamental importance.
At a sufficiently low temperature,
most of classical systems show some ordered structure 
which implies that the symmetry 
at the high temperature is spontaneously lost or reduced.
This is the spontaneous symmetry breaking which 
 is usually characterized by using 
a {\em local } order parameter as an existence of the long range order.
States of matter in a classical system are mostly characterized by 
this order parameter with the symmetry breaking.
Even in a quantum system, the local order parameter and
the symmetry breaking play similar roles
and they form a foundation of our physical understanding. 
Typical examples can be ferromagnetic and Neel orders in spin systems.

Recent studies in decades have revealed that
this symmetry breaking may not be always enough to 
characterize some of important quantum states\cite{wen89,Hatsugai04e}.
Low dimensionality of the system
 and/or geometrical frustrations come from the strong correlation
can prevent from a formation of the local order.
Especially with a quantum fluctuation,
 there may happen that a quantum ground state without
any explicit symmetry breaking is realized 
even in the zero temperature.
Such a state is classified as a quantum liquid
which mostly has an energy gap (may not be always).
Typical example of this quantum liquids
is the Haldane spin chain and the valence bond solid (VBS) 
states\cite{Haldane83-c,Affleck87-AKLT}.
Also some of the frustrated spin systems and 
spin-Peierls systems can belong to this 
class\cite{Rokhasar88,Read91-LN,Sondi01}. 
 To characterize these quantum liquids,
a concept of a topological order can be useful\cite{wen89,Hatsugai04e}.
It was proposed to characterize quantum Hall states
which are typical quantum liquids with energy gaps.
There are many clearly different quantum states but 
they do not have any 
local order parameter associated with symmetry breaking.
Then topological quantities such as a number  of degenerate 
ground states and the Chern numbers as the Hall conductance
are used to characterize the quantum liquids.
We generalize the idea to use the topological quantities
such as the Chern numbers for the characterization 
of the generic quantum liquids\cite{Hatsugai04e}.
This is a global characterization. 
When we apply this to spin systems with 
the time-reversal symmetry (TR), the Chern number
is vanishing in most cases.
Recently we propose an alternative for the system
with the TR invariance by the quantized Berry phases\cite{Hatsugai06a}.
Although, the Berry phases 
can take any values generically, 
the TR invariance of the ground state
guarantees a quantization of the Berry phases
which enables us to use them as  local topological order 
parameters.
In the present article, we use it 
for several spin systems with frustrations and verify the validity.
Although the geometrical frustration affects the 
standard local order 
substantially, 
it does not bring any fundamental difficulties
for the topological characterizations
as shown later.
It should be quite useful for characterizations
for general quantum liquids\cite{Hatsugai06a}.

Finally we mention on the energy spectra of
the systems with classical or topological orders.
There can be interesting differences between 
the standard order and the topological order.
As for energy spectra, we have two situations when the
symmetry is spontaneously broken.
If the spontaneously broken symmetry  is continuous, 
there exists
a gapless excitation as a Nambu-Goldstone mode.
On the other hand, the symmetry is discrete, the ground
states are  degenerate and above these degenerate states,
 there is a finite energy gap. 
Note that when 
the system is finite (with periodic boundary condition),
the degeneracy is lifted by the small energy gap, $e^{-L^d/\xi}$,
where $L$, $d$ and $\xi$ are a linear dimension of the finite system,
dimensionality  and a typical correlation length.
For the topological ordered states with energy gaps, 
we may expect degeneracy of the ground states depending on the
geometry of the system (topological degeneracy).
When the system is finite, we expect edge states
generically\cite{Hatsugai93b}. It implies the topological degeneracy is lifted by
 the energy gaps of the order $e^{-L/\xi}$.

\section{Local Order Parameters of Quantum Liquids}
After the first discovery of the 
fractional quantum Hall states, 
the quantum liquids have been recognized to exist
quite universally in a quantum world where 
quantum effects 
can not be treated as a correction to the classical description
and the quantum law itself takes the wheel to determine 
the ground state. 
The resonating valence bond (RVB) state
which is proposed for a basic platform of 
the high-$T_C$ superconductivity
is a typical example\cite{Anderson87}.
The RVB state of the Anderson can be understood as 
a quantum mechanical collection of {\em local} spin singlets.
When it becomes mobile under the doping,
the state is expected to show superconductivity.
Original ideas of this RVB  go back to the Pauling's
description of benzene compounds where 
the quantum mechanical ground state is composed of 
{\em local bonding states (covalent bonds) }
where the basic variables to describe the state
is not electrons localized at sites but 
the bonding states on  links\cite{Pauling}.
This is quite instructive. That is, 
in both  of 
the Anderson's RVB and the Pauling's RVB,
basic objects to describe the quantum liquids are 
 quantum mechanical objects as a {\em singlet pair}
and a {\em covalent bond}\cite{Hatsugai06a}.
The ``classical'' objects as small magnets (localized spins)
and electrons at site never play  major roles.
The  constituents of the liquids themselves
do not have a classical analogue and 
purely quantum mechanical objects.
Based on this view point, 
it is natural to characterize these quantum objects,
the singlet pairs and the covalent bonds, as working variables of
the  {\em local} quantum order parameters.
It is to be compared with the conventional order parameter
(a magnetic order parameter is defined by a local spin as a working variable).
From these observations,
we proposed to use quantized Berry phases 
to define local  topological order parameters\cite{Hatsugai06a}. 
( We only treat here the singlet pairs as the topological order parameters. As for
the local topological description by the covalent bonds, see ref.[1].) 
For example, 
there can be many kinds of quantum dimer states for frustrated 
Heisenberg models, such as
column dimers, plaquette dimers, etc. 
As is clear, one can not find any classical local order parameters
to characterize them.
However, our topological order parameters can distinguish
them as different phases not by just a crossover. 

\section{Quantized Berry Phases for the Topological Order Parameters
of Frustrated Heisenberg Spins}
Frustration among spins prevent from forming a magnetic order
and their quantum ground states tend to belong to the quantum liquids
without any symmetry breaking. 
Since they do not have any local order parameters,
even if they have apparent different physical behaviors,
 it is difficult to
make a clear 
distinction as a phase
not just as a crossover.
We apply the general scheme in the reference [1] 
to classify these frustrated spin systems.
Defining quantized Berry phases as $0$ or $\pi$,
the spin liquids are characterized locally reflecting their topological order.
We can distinguish many topological phases which are separated by 
 local quantum phase transitions (local gap closings).

We consider following  spin $1/2$ Heisenberg models with general exchange couplings,
$
H
=  \sum_{ij}
{J }_{ij}{\mb{S} _i} \cdot
\mb{S} _j 
$.
{\em  We  allow  frustrations among spins. }
We assume the ground state is {\em unique and gapped}.
To define a local topological order parameter
at a specific link $ \langle ij \rangle  $, 
 we modify the exchange 
by making a local $SU(2)$ twist $\theta $ only at the link as 
\begin{eqnarray*}
 J_{ij}\mb{S} _i \cdot \mb{S} _j
&\to & 
J_{ij}
\big(
 \frac {1}{2} 
(
e^{- i\theta }
S_{i+} S_{j-}
+
e^{ i \theta}
S_{i-}S_{j+}
)
+  S_{iz}S_{jz}
\big).
\end{eqnarray*} 
Writing $x=e^{i\theta}$, we define a 
parameter dependent Hamiltonian $H(x)$ and its normalized 
ground state $|\psi(x) \rangle $ as
$H(x) |\psi(x) \rangle =E(x) | \psi(x) \rangle $, 
$\langle {\psi} | {\psi} \rangle= 1$.
Note that this Hamiltonian is invariant under the time-reversal (TR) $\Theta_T$,
$
 \Theta_{ T} ^{-1} 
H(x)
\Theta_{ T} 
=  H(x)
$\cite{tri}.
Also note that by 
changing $\theta:0\to 2\pi$, 
we define a closed loop $C$ in the parameter space of $x$.

Now we define the Berry connection as 
$
{A}_\psi  = \langle {\psi} |  d {\psi} \rangle  = 
 \langle {\psi} |  \frac {d  }{d  x}  \psi\rangle   dx
$. Then the Berry phase along the loop $C$ is defined
as $
i{\gamma } _C ({A}_\psi )= \int_C {A}_\psi 
$\cite{berry84}. 
Besides that the system is gapped, 
we further assume {\em the excitation gap is always finite} (for $^\forall x$),
to ensure the regularity of the ground state\cite{Hatsugai04e}.
This may not be alway true, since  the gap can collapse by the local perturbation
as an appearance of localized states (edge states)\cite{Hatsugai93b}.
Note that by changing a  phase of the ground state as 
$| {\psi}(x) \rangle =| {\psi}^\prime(x) \rangle 
e^{i\Omega(x)} $, 
the Berry connection gets modified as 
$A_\psi= {A}_\psi^\prime   + i d {\Omega}  $
\cite{berry84,Hatsugai04e}.
It is a gauge transformation.
Then the Berry phase, $\mb{\gamma }_C $ also changes.
It implies that the Berry phase
 is not well defined without specifying the phase of
 the ground state (the gauge fixing).
It can be 
 fixed by taking a single-valued reference state $|\phi \rangle $ and
a  gauge invariant  projection into the ground state 
$ 
P =
|  \psi \rangle \langle  \psi |=
|  \psi^\prime \rangle  \langle  \psi^\prime|
$ as 
$
|{\psi}_\phi  \rangle =  {P} |{\phi} \rangle  /\sqrt{N_\phi }$,
$N_\phi = \|{P} |\phi \rangle \|^2
= |\eta_\phi| ^2$,
$\eta_\phi= \langle \psi | \phi  \rangle $\cite{Hatsugai04e,Hatsugai06a}.
We here require 
the normalization, $N_\phi$, 
to be finite.
When we use another reference state $| \phi^\prime \rangle  $ to fix the gauge,
we have 
$
|\psi_\phi  \rangle =
|  \psi_{\phi^\prime} \rangle 
e^{i \Omega },\ 
{\Omega} =
{\rm arg}\, (
{\eta}_\phi  - {\eta}_{\phi'} )
$.
Due to this gauge transformation, 
the Berry phase gets modified as 
$ {\gamma } _{C} ({A}_{\psi_\phi} ) = 
{\gamma } _{C} ({A}_{\psi_{\phi^\prime}} )+ \Delta, \quad
\Delta_{}=    \int_C 
  d {\Omega} $. 
Since the reference states $|\phi \rangle $ and  $|\phi' \rangle $ are
 single-valued on the $C$,
the phase difference $\Omega $
is just different by 
 $\Delta=2\pi  M_C   $ with some integer $M_C$.
Generically it implies that the Berry phase 
 has a gauge invariant meaning just up to the integer as 
\begin{eqnarray*}
\gamma _C & \equiv &
-i \int_C\,{A},\quad {\rm mod}\, 2\pi 
\end{eqnarray*} 

By the TR invariance, 
the  Berry phase get modified as
$
\gamma _C (A_\psi) = \sum_J C_J^* d C_j=-\sum_J C_J d C_j^*=
  -\gamma _C(A_{\Theta\psi})
$ since $\sum_J|C_J|^2=1$\cite{Hatsugai06a}.
Therefore to be compatible with
the gauge ambiguity,
the Berry phase of the unique TR-invariant ground state, 
$|\psi \rangle \propto \Theta| \psi \rangle $,
 satisfies $\gamma _C (A_\psi) \equiv -\gamma _C(A_{\psi})\ ( {\rm mod}\, 2\pi)$.
Then it is required to
{\em be quantized}
as
\begin{eqnarray*}
\gamma _C(A_\psi) &=& 0, \pi \ ({\rm mod}\ 2\pi ).
\end{eqnarray*} 
This quantized Berry phases have a topological stability
since any small perturbations can not modify
unless the gauge becomes singular.
Here we note that 
the Berry phase of the singlet pair for the two site problem is 
 $\pi$\cite{Hatsugai06a}. 
Now let us  take any dimer covering of all sites 
 ${\cal D}=\{\langle ij \rangle  \}$
($\#{\cal D}=N/2$, $N$ is a total number od sites) and 
assume that the interaction is  nonzero only on these dimer links,
then the Berry phases, $\pi$,
 pickup the dimer pattern $\cal D$.
Now imagine an adiabatic process to include interactions across
 the dimers.
Due to the topological stability of the quantized Berry phase,
they can not be modified unless the dimer gap collapses.
This dimer limit presents 
a non-trivial pattern of a quantized Berry phase 
and shows the usefulness
of the quantized Berry phases as {\em local order parameters 
of singlet pairs}. 
To show its real validity of the quantized Berry phases,
we have diagonalized the Heisenberg Hamiltonians numerically by
the Lanzcos algorithm and calculated the quantized Berry phases
explicitly.

\begin{figure}
\begin{center}
\includegraphics[width=1.0\linewidth,clip]{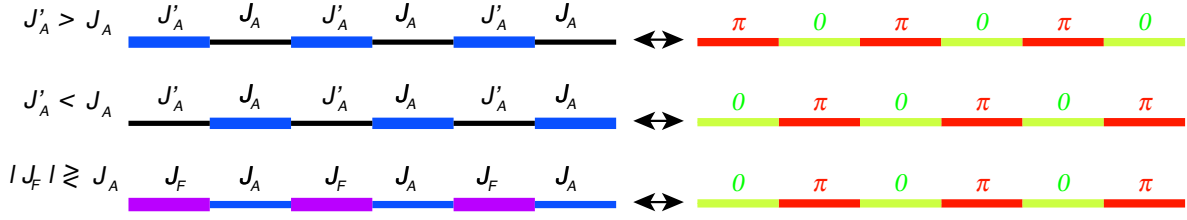}%
\end{center}
 \caption{
One dimensional Heisenberg models with
alternating exchange interactions
with periodic boundary condition (left).
Numerically evaluated distribution of the quantized Berry phases (right).
$J_A, J_{A'}>0$ and $J_F<0$. 
The results are independent of the system size.
( We have checked a consistency of the results for various possible system sizes.)
\label{f:1D}
}
\end{figure}

The first numerical examples  are the Heisenberg chains with alternating exchanges.
When the exchanges are both antiferromagnetic as $J_A>0$ and  $J_{A'}>0$,
it is a spin Pierls or dimerized chain.
In this case, the Berry phases are $\pi$ on the links with the strong exchange
couplings and $0$ on the one with the weak couplings (Fig.\ref{f:1D}).
This is expected from the adiabatic principle and the quantization.
When one of them is negative as $J_A>0$ and  $J_{F}<0$,
the calculated Berry phases are $\pi$ for the antiferromagnetic links
and $0$ for the ferromagnetic ones. It is independent of the
ratio $J_A/J_F$.
Since the strong ferromagnetic limit is equivalent with the spin $1$ chain,
 it is consistent with the topological nontrivial structure
of the Haldane phases. Further analysis on the $S=1$ systems will be published 
elsewhere.
\begin{figure}
\begin{center}
\includegraphics[width=1.0\linewidth,clip]{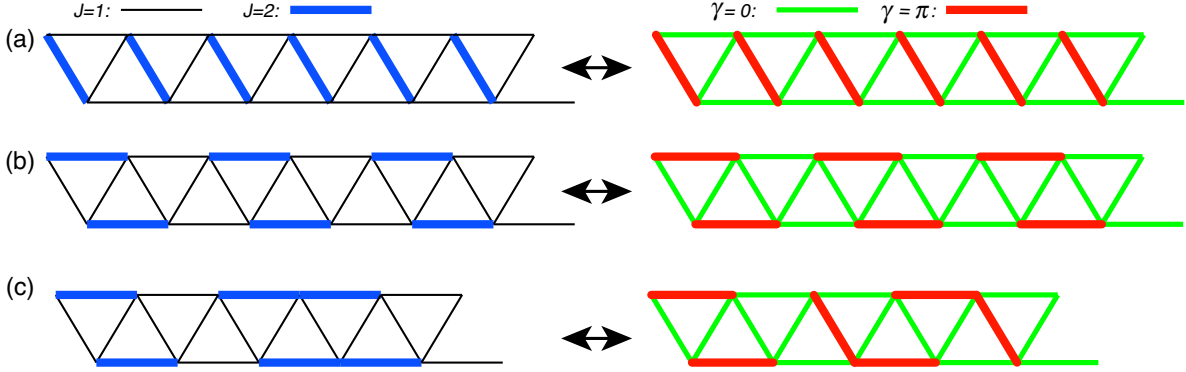}%
\end{center}
 \caption{
One dimensional Heisenberg models with NN and NNN exchanges (left) 
with periodic boundary condition.
Numerically evaluated distribution of the quantized Berry phases (right).
(a), (b) and (c): three different exchange configurations of $J=1$ and $ J'=2$.
\label{f:tri}
}\vskip -0.6cm
\end{figure}
Next  numerical examples are spin chains with nearest neighbor (NN) 
and next nearest neighbor (NNN) exchanges
as  ladder of triangles (Fig.\ref{f:tri}).
These are typical systems with frustrations.
(a) and (b) are two different but specific configurations
where one may adiabatically connect the system with
different dimer coverings by the strong coupling bonds. 
In these cases, the quantized Berry phases are $\pi$
for the strong coupling links and $0$ for the rest links.
This is consistent with the adiabatic principle. 
We note here that 
it is difficult to make a qualitative difference 
between the two quantum liquids by a conventional methods.
 However we have
made a clear distinction between them as two different 
topological phases.
The present scheme is not only valid for these simple situations but also
useful for generic situation. 
For example, as for a system in the Fig.\ref{f:tri} (c),
 we can not use the adiabatic principle simply.
However the quantized Berry phases show non trivial behaviors and
it make a clear distinction  
 that the phase (c) is topologically  different from the ones in the 
(a) and (b) 
as an independent phase not just as a crossover. 
A local quantum phase transition separates them by the gap closing.
As is now clear, the present scheme is quite powerful to make
a local characterization of the
topological quantum insulators.

Part of the  work  was supported by 
Grant-in-Aids for Scientific Research
 (Grant No. 17540347) and 
on Priority Areas
(Grant No. 18043007) from MEXT,
and the Sumitomo Foundation.



\begin{thebibliography}{99}
\expandafter\ifx\csname natexlab\endcsname\relax\def\natexlab#1{#1}\fi
\expandafter\ifx\csname bibnamefont\endcsname\relax
  \def\bibnamefont#1{#1}\fi
\expandafter\ifx\csname bibfnamefont\endcsname\relax
  \def\bibfnamefont#1{#1}\fi
\expandafter\ifx\csname citenamefont\endcsname\relax
  \def\citenamefont#1{#1}\fi
\expandafter\ifx\csname url\endcsname\relax
  \def\url#1{\texttt{#1}}\fi
\expandafter\ifx\csname urlprefix\endcsname\relax\def\urlprefix{URL }\fi
\providecommand{\bibinfo}[2]{#2}
\providecommand{\eprint}[2][]{\url{#2}}

\bibitem{Hatsugai06a}
Y. Hatsugai, preprint, cond-mat/0603230.

\bibitem{wen89}
\bibinfo{author}{\bibfnamefont{X.~G.} \bibnamefont{Wen}},
  \bibinfo{journal}{Phys. \ Rev.\ B} \textbf{\bibinfo{volume}{40}},
  \bibinfo{pages}{7387} (\bibinfo{year}{1989}).


\bibitem{Hatsugai04e}
\bibinfo{author}{\bibfnamefont{Y.}~\bibnamefont{Hatsugai}},
  \bibinfo{journal}{J.\ Phys. \ Soc.\ Jpn.} \textbf{\bibinfo{volume}{73}},
  \bibinfo{pages}{2604} (\bibinfo{year}{2004}),
 \textbf{\bibinfo{volume}{74}},
  \bibinfo{pages}{1374} (\bibinfo{year}{2005}).


\bibitem{Haldane83-c}
\bibinfo{author}{\bibfnamefont{F.~D.~M.} \bibnamefont{Haldane}},
  \bibinfo{journal}{Phys.\ Lett.} \textbf{\bibinfo{volume}{A93}},
  \bibinfo{pages}{464} (\bibinfo{year}{1983}).

\bibitem{Affleck87-AKLT}
\bibinfo{author}{\bibfnamefont{I.}~\bibnamefont{Affleck}},
  \bibinfo{author}{\bibfnamefont{T.}~\bibnamefont{Kennedy}},
  \bibinfo{author}{\bibfnamefont{E.~H.} \bibnamefont{Lieb}}, \bibnamefont{and}
  \bibinfo{author}{\bibfnamefont{H.}~\bibnamefont{Tasaki}},
  \bibinfo{journal}{Phys.\ Rev.\ Lett.} \textbf{\bibinfo{volume}{59}},
  \bibinfo{pages}{799} (\bibinfo{year}{1987}).



\bibitem{Rokhasar88}
\bibinfo{author}{\bibfnamefont{D.~S.} \bibnamefont{Rokhsar}} \bibnamefont{and}
  \bibinfo{author}{\bibfnamefont{S.~A.} \bibnamefont{Kivelson}},
  \bibinfo{journal}{Phys.\ Rev.\ Lett.} \textbf{\bibinfo{volume}{61}},
  \bibinfo{pages}{2376} (\bibinfo{year}{1988}).

\bibitem{Read91-LN}
\bibinfo{author}{\bibfnamefont{N.}~\bibnamefont{Read}} \bibnamefont{and}
  \bibinfo{author}{\bibfnamefont{S.}~\bibnamefont{Sachdev}},
  \bibinfo{journal}{Phys.\ Rev.\ Lett.} 
 \textbf{\bibinfo{volume}{62}},
  \bibinfo{pages}{1694} (\bibinfo{year}{1989}).


\bibitem{Sondi01}
\bibinfo{author}{\bibfnamefont{R.}~\bibnamefont{Moessner}} \bibnamefont{and}
  \bibinfo{author}{\bibfnamefont{S.~L.} \bibnamefont{Sondhi}},
  \bibinfo{journal}{Phys.\ Rev.\ Lett.} \textbf{\bibinfo{volume}{86}},
  \bibinfo{pages}{1881} (\bibinfo{year}{1989}).

\bibitem{Anderson87}
\bibinfo{author}{\bibfnamefont{P.~W.} \bibnamefont{Anderson}},
  \bibinfo{journal}{Science} \textbf{\bibinfo{volume}{235}},
  \bibinfo{pages}{1196} (\bibinfo{year}{1987}).

\bibitem{Pauling}
L. Pauling, Proc. Nat. Acad. Sci. 39, 551 (1953).


\bibitem{tri}
The TR is defined as 
 $\Theta_T=K \otimes_j(i \sigma _{jy})$, 
as the  anti-unitary operation ($K$: complex conjugation). 
It operates for a state 
$
|G \rangle = \sum_{J=\{\sigma _1,\cdots,\sigma_N\}}
 C_J | \sigma _1,\sigma _2,\cdots,\sigma _N \rangle
$,
($\sigma _i=\pm 1$)
as 
$
\Theta_{ T} |G \rangle = \sum C_J^* (-)^{\sum_{i=1}^N (1+\sigma _i)/2}
| -\sigma _1,\cdots,-\sigma _N \rangle 
$.
Then spins get transformed as
$
^\forall j,\ \mb{S}_j \to  \Theta_{ T} ^{-1}  \mb{S}_j \Theta_{ T} = - \mb{S} _j
$ and $\mb{S}_i\cdot \mb{S}_j $ is a TR invariant.

\bibitem{berry84}
\bibinfo{author}{\bibfnamefont{M.~V.} \bibnamefont{Berry}},
  \bibinfo{journal}{Proc.\ R.\ Soc.} \textbf{\bibinfo{volume}{A392}},
  \bibinfo{pages}{45} (\bibinfo{year}{1984}).



\bibitem{Hatsugai93b}
Y. Hatsugai,
Phys. \ Rev.\ Lett. 
\textbf{\bibinfo{volume}{71}}, 3697 (1993),

\end{thebibliography}
\end{document}